\def \ba {\begin{array}}
\def \ea {\end{array}}

\def \bea {\begin{eqnarray}}
\def \eea {\end{eqnarray}}

\def \be {\begin{equation}}
\def \ee {\end{equation}}

\def\nn{\nonumber}

\documentclass[11pt,a4]{article}
\usepackage{amssymb,amsfonts}
\begin{document}

\begin{center} 
{\bf Space-time dynamics from algebra representations}
\footnote{Work partially
supported by the DGICYT.}
\end{center}
\bigskip
\bigskip
\centerline{ {\it J.L. Jaramillo$^{1}$\footnote{E-mail: jarama@iaa.es} and  
 V. Aldaya$^{1,2}$\footnote{E-mail: valdaya@iaa.es}} }
\bigskip

\begin{enumerate}
\item {Instituto de Astrof\'{\i}sica de Andaluc\'{\i}a (CSIC), Apartado Postal 3004,
18080 Granada, Spain.}
\item  {Instituto Carlos I de F\'\i sica Te\'orica y Computacional, Facultad
de Ciencias, Universidad de Granada, Campus de Fuentenueva, 
Granada 18002, Spain.} 
\end{enumerate}

\bigskip
\begin{center}
{\bf Abstract}
\end{center}
\small

\begin{list}{}{\setlength{\leftmargin}{3pc}\setlength{\rightmargin}{3pc}}
\item 
We present a model for introducing dynamics into a space-time geometry. This 
space-time structure is constructed from a $C^*$-algebra defined in terms of
the generators of an irreducible unitary representation of a finite-dimensional
Lie algebra ${\cal G}$. This algebra is included as a subalgebra in a bigger 
algebra ${\cal F}$, the
generators of which mix the representations of ${\cal G}$ in a way that 
relates different
space-times and creates the dynamics. This construction can be considered 
eventually as a model for
$2-D$ quantum gravity.

\end{list}
03.65.Fd, 98.80.H, 02.20.Sv, 02.20.Tw.
\normalsize

\vskip 1cm

A $C^*$-algebra constructed by means of the generators of a finite-dimensional
Lie algebra ${\cal G}$ is employed to construct an appropriate space-time 
geometry
structure from strictly symmetry grounds. Moreover, a notion of dynamics
(in a sense to be specified below) can be introduced into the scheme by 
considering ${\cal G}$ as inserted in a bigger algebra ${\cal F}$ which 
encloses the physical content of the model.
The method somewhat parallels that of Madore in constructing the {\it fuzzy 
sphere} \cite{Madore}. While he employed the generators in irreducible unitary
representations of $su(2)$, we make use here of the algebra of 
$sl(2,{\mathbb R})$ to
create what might be called 
{\it fuzzy hyperboloids} (there are different ways of introducing 
non-commutativity in hyperboloids, thus leading to different ``fuzzy
hyperboloids''; we shall give a precise meaning to {\it ours} below) .

For the sake of clarity and to avoid redundancies, let us postpone the
detailed analysis of the concrete $C^*$-algebra of interest until we have 
chosen a specific model in which our statements acquire a completely defined
meaning. For the moment let us accept that a geometry notion (an hyperboloid)
can be related
to each particular irreducible representation of ${\cal G}$. 

To introduce dynamics into this context, we consider a bigger algebra 
${\cal F}$ with a central extension structure. This algebra contains
$sl(2,{\mathbb R})$ as a subalgebra singularized by algebra
(pseudo-)cohomology criteria: $sl(2,{\mathbb R})$ is in the kernel of the 
cocycle. Generators which give a central term in their 
commutators are 
dynamical and form conjugated pairs, while those in the 
kernel of the cocycle are the kinematical ones (this can
be explicitly shown through the construction of a symplectic form).
Then, we construct a unitary, 
irreducible representation of ${\cal F}$. Assuming complete reducibility of the
representation under $sl(2,{\mathbb R})$, we find a 
collection of $sl(2,{\mathbb R})$ irreducible, unitary representations, each  
defining a space-time geometry, via the $sl(2,{\mathbb R})$ generators. The action of
the rest of the generators in ${\cal F}$ mix the different $sl(2,{\mathbb R})$ 
representations, defining dynamics in the ensemble of the {\it hyperboloids}.

Let us make the foregoing considerations more specific. For $2D$ quantum 
gravity motivations \cite{Polyakov,Alekseev,Miguel}, we choose ${\cal F}$ 
to be the Virasoro algebra:
\be
[ L_n,L_m] =( n-m) L_{n+m}+\frac 1{12}(
cn^3-c'n) \delta _{n,-m} , \label{conmutacion}
\ee
where $c$ is the true extension parameter and $c'$ is a redefinition of 
$L_0$ (pseudo-cohomology), which must be taken into account in order to
fully explore the dynamical content of the algebra \cite{Julio} 
(the standard expression
in the literature for (\ref{conmutacion}) uses $c$ and $h$ 
instead of $c$ and $c'$, where $h=\frac{c-c'}{24}$; parameters $(c,c')$ are
better suited for our discussion).

Then, a highest-weight representation of the Virasoro algebra, 
${\cal H}_{(c,c')}$, is constructed. 
Unitarity, irreducibility \cite{Kac,Feigin&Fuchs,Thorn,Senechal,Navarro-Salas}
and the presence of $sl(2,{\mathbb R})$ in the kernel of the cocycle 
(i.e., $sl(2;R)$ is kinematical) are automatically guaranteed with the 
imposition of $c=c'$ and $c>1$. This 
$sl(2,{\mathbb R})$ is generated by $\langle L_1, L_0, L_{-1}\rangle$.

The representation ${\cal H}_{(c,c)}$ is accomplished by imposing 
$L_n\mid 0\rangle =0$ (for $n\geq -1$, annihilation operators), and the 
states have the form:
\be
L_{n_1}...L_{n_j}\mid 0\rangle \; \; \; (n_1,...,n_j \leq 2 \hbox{ creation 
operators)}
\ee

To consider the reduction of the representation under the kinematical
subalgebra $sl(2,{\mathbb R})$, we look for $\mid N,0 \rangle$ states,
satisfying $L_1\mid N,0 \rangle=0$, thus being highest-weight 
vectors for $sl(2,{\mathbb R})$. It can be shown \cite{QG2} 
that each of these vectors must belong to a definite Virasoro level (i.e., 
they must be $L_0$-eigenvectors). Using this fact, and denoting as $D^{(N)}$ 
the dimension of the Virasoro $N$ level, we find $(D^{(N)}-D^{(N-1)})$ 
highest-weight
vectors in the Virasoro $N$ level (we need only notice that the operator
$L_1$ restricted to $Level(N)$ with values in $Level(N-1)$ is an epimorphism, and $dim(Level N) =
dim(Ker L_1)+dim(Im L_1)$). An $sl(2,{\mathbb R})$ representation $R^{(N)}$, of
Casimir value $N(N-1)$, is reached by the successive action of $L_{-1}$ 
on each of the previously found
$\mid N,0\rangle$ vectors: $\mid N,n\rangle=(L_{-1})^n \mid N,0\rangle$. Furthermore,
the different $sl(2,{\mathbb R})$ representations are orthogonal (with the scalar
product induced from the Virasoro algebra). The representation of the
Virasoro algebra is thus completely reduced:
\be
{\cal H}_{(c,c)}=\bigoplus (D^{(N)}-D^{(N-1)})R^{(N)}
\ee

Note that the representation $R^{(N)}$ is degenerated and its weight 
$(D^{(N)}-D^{(N-1)})$ increases with $N$.

As stated above, the space-time is reconstructed from a $C^*$-algebra,
and in the search of it we follow the spirit of Madore 
\cite{Madore} in the realization of the ``fuzzy sphere'', but with a 
different objective which results in a different construction.

The aim in \cite{Madore} was to construct a non-commutative geometry for the 
sphere in such a way that the classical 
geometry is recovered in a certain limit. This was achieved through 
the construction of a 
succession of $C^*$-algebras the limit of which is the algebra 
${\cal C}(S^2)$ of complex-valued smooth functions on the sphere.

The explicit sphere was defined by:\\
\be
g_{ab}x^ax^b=r^2 \hbox{   (}r\hbox{ being a fixed radius)} \label{radius} \\
\ee
$g_{ab}$ being the Killing metric.

To implement the $n-th$ element of the succession of $C^*$-algebras, he 
defined 
``coordinates'' from the $J_{n}^a$ generators of an irreducible 
representation of 
dimension $n$ of $su(2)$:
\be
{x}_{n}^{a}=k_n {J}_{n}^{a} \hbox{   ,(where } k_n \hbox{ is a constant with 
appropriate dimensions).} 
\ee
When polynomials were considered in these non-commutative coordinates of 
order up to $n-1$, with the
Casimir constraint (\ref{radius}), an algebra isomorphic to
$M_n$ ($n\times n$ matrices) resulted. This non-commutative $C^*$ algebra 
$M_n$ was then 
used to construct a matrix geometry which in the limit $n\rightarrow\infty$ 
goes to the standard geometry on the sphere of radius $r$.

Geometry becomes fuzzy in this process. For each matrix geometry $M_n$, 
points are replaced by states of the n-dimensional $su(2)$ 
representation considered. We can prove that $k\rightarrow 0$ in the limit 
$n\rightarrow\infty$, and thus coordinates become commutative, allowing the 
characterization of a point by the use of two coordinates (recovering the 
standard notion of a point).

In our case, the starting algebra is $sl(2,{\mathbb R})$ instead of $su(2)$, so that 
{\it hyperboloids} substitute spheres
when the Casimir constraint is imposed.

$C^*$-algebras are again built from the representations of our Lie algebra
(and thus we are in the spirit of Madore), but now we are not trying to 
approximate any previous classical geometry (true hyperboloids of ``radii''
$r$),
and thus a succession of representations of the algebra for implementing
such an approximation is not required. In our case, there is no arbitrariness in the $sl(2,{\mathbb R})$ 
representations we must consider. They are specific ones and are given by the 
reduction of the ${\cal H}_{(c,c)}$ representation. Each $sl(2,{\mathbb R})$ 
representation, $R^{(N)}_i$ ($i$ for degenerate representations), will 
generate a (different) space-time geometry. 

To construct the $C^*$-algebras, we first define the coordinate
variables from the generators of $sl(2,{\mathbb R})$. 
Generators in the $N-th$  $sl(2,{\mathbb R})$ representation are multiplied by an 
dimensional constant $k^{(N)}$ in order to get appropriate space-time 
coordinates:\\
\bea 
x_{i}^{(N)}&=&k^{(N)} L_{i}^{(N)} \hbox{ \ \ }  i=-1,0,1\\
\hbox{where  } L_{i}^{(N)}&=&L_{i} \left|_{N-th sl(2,{\mathbb R}) \ \
representation  }\right.
\hbox{.} \nn 
\eea  
We impose the condition that
all the {\it hyperboloids} derived from the different $sl(2,{\mathbb R})$ 
representations in the Virasoro representation have
the same {\it radius}, $R$, and this fixes the value of the 
constants $k^{(N)}$.
The way
the radius is implemented in an $sl(2,{\mathbb R})$ irreducible representation is, 
again, via the value of the Casimir on it (in fact, we have imposed 
irreducibility on these representations in order to have a well defined value 
of the Casimir):
\be
-R^2=g^{jk}x_{j}^{(N)} x_{k}^{(N)}={k^{(N)}}^2 N(N-1)\hbox{,}
\ee
where $g^{jk}$ is the $sl(2,{\mathbb R})$ Killing metric (note the condition 
$\frac{{k^{(N)}}^2}{\mid{k^{(N)}}^2\mid}=-1$, i.e., $k^{(N)}$ is a purely 
imaginary number).

Thus, we finally have:
\bea
R^2&=&-{k^{(N)}}^2 N(N-1)\\
k^{(N)}&=&i\frac{R}{\sqrt{N(N-1)}} . \nn
\eea
This is the way space-time variables are defined. To implement 
the 
$C^*$-algebra, we do not restrict ourselves to polynomials up to a certain 
order 
(we are not trying to define a sequence of space-times), but rather, we consider the 
entire enveloping algebra
of these $x_{i}^{(N)}$, modulus the ideal generated by the Casimir (radius)
constraint. 
Thus,
\bea
C^*-algebra&=&Env(\langle x_{-1}^{(i)},x_{0}^{(i)},x_{1}^{(i)} \rangle)/Radius \nn  \\
Radius&=& -g^{jk}x_{j}^{(N)} x_{k}^{(N)}= R^2 .
\eea
\\
As can be seen from commutators among the space-time coordinates ($[x_{i}^{(N)}
\hbox{,}x_{j}^{(N)}]=k^{(N)}C_{ij}^k x_{k}^{(N)}$), this is a non-commutative
$C^*$-algebra leading to a non-commutative 
geometry. Points are again replaced by states in the representation of 
$sl(2,{\mathbb R})$, and thus we have indeed {\it fuzzy hyperboloids}. For a
better understanding of these ``fuzzy'' points, it is useful a glance at the 
indetermination relations, under which space-time is divided into cells:
\be
\Delta x_{i}^{(N)}\Delta x_{j}^{(N)}\geq {\mid k^{(N)}\mid }^2=\frac{R^2}{\sqrt{N(N-1)}} . \label{indet}
\ee
Different fuzzy hyperboloids of the same radius $R$ are simultaneously found 
inside the Virasoro 
representation. These are distinguished by point density, which grows with
the value of $N$, as can be seen from (\ref{indet}). We note that for large 
$R$ values and very small $N$, the size of the cells is comparable to that 
of the hyperboloid. On the contrary, for a fixed $R$, we can find values of 
$N$ as large as we wish, making 
$k^{(N)}\rightarrow 0$, so that cells tend to 
points and the space-time coordinates become commutative, {\it recovering} the 
classical geometry. 

Our model for space-time is not just one of these hyperboloids, but the whole
ensemble of them (they can be seen as different copies of the same hyperboloid,
with equal $R$, but with different degrees of {\it fuzziness}, different $N$).
We understand ``point'' to mean a normalized state in the Virasoro Hilbert 
space 
${\cal H}_{(c,c)}$. Taking advantage of the complete reduction of 
${\cal H}_{(c,c)}$ under $sl(2,{\mathbb R})$, this point can be written as a linear
superposition of normalized vectors over the $sl(2,{\mathbb R})$ representations. Each of
$sl(2,{\mathbb R})$ states is interpreted as a ``point'' in a concrete {\it fuzzy 
hyperboloid}, 
and thus the original point is spread over different hyperboloids. Indeed,
it makes sense to consider the probability of the ``point'' to be in a 
concrete hyperboloid using the orthogonality of the $sl(2,{\mathbb R})$ 
representations and developing a standard quantum mechanical interpretation.

It is not our aim here to give a detailed analysis of the {\it fuzzy 
hyperboloid} geometry and its classical (large $N$) limit. We simply 
mention general 
features.
The role of space-time 
diffeomorphisms is played by automorphisms of the $C^*$-algebra. 
Vector fields are derivations of this algebra; that is, linear mappings that
satisfy the Leibnitz rule. These fields do not form a module over $C^*$,
suggesting that we should  
consider one-forms as the fundamental objects having a bimodule
structure over $C^*$ \cite{Connes}. From this, and the Killing metric on 
$sl(2,{\mathbb R})$, we could
even define a metric and a connection (we do not enter into these details, 
which are subtle and deserve a specific study; basically, we aim to identify a 
proper $C^*$-algebra).

Let us focus now on the way the dynamical degrees of freedom enter the 
model. Since the 
motivation for the use of the Virasoro algebra is $2-D$ gravity, we shall
refer to these modes as ``gravitational'' ones. Their action on points 
(normalized
states in the Virasoro Hilbert space) must be such that it preserves the norm 
(keeping
the notion of ``point''). Therefore, they must be implemented by unitary 
transformations 
generated by hermitian operators. Starting from the condition\\
\be
L_n^+=L_{-n} ,
\ee
hermitian combinations can be defined:\\
\bea
G_n &=& L_n + L_{-n} \; \hbox{,} \; n\geq 2 \\
G_{-n} &=& i(L_n - L_{-n}) \; \hbox{,} \; n\geq 2 , \nn
\eea
generating the unitary gravity transformations:\\
\be
U_n=e^{i(k_n G_n + k_{-n} G_{-n})}\hbox{  .} 
\ee
Gravity transformations do not preserve the $sl(2,{\mathbb R})$ 
irreducible representations, so that if we start from a point 
which completely lies on a  
space-time of point density given by $N$, after the action of gravity this
point is transformed  into a superposition of points in different 
space-times of the same radius (the same Virasoro representation) but 
different $N$ (different point density). This is the essence of the dynamics 
in the model: the Universe
is not one of these space-times, but the whole ensemble of them, and a point 
is a superposition of states (eigen-points) spread over different-density 
space-times, the
coefficients of which give 
the probability for the point to be in
the corresponding space-time (because $sl(2,{\mathbb R})$ representations are 
orthogonal, a fact which allows the construction of proper orthogonal 
projectors), thereby 
defining a probability distribution of the point. 
The effect of (gravity) dynamics is that of changing
this probability distribution (quantum motion of the point).\\

Space-times with different densities have different weights, in such a way that
denser ones (more ``classical'' ones) are more abundant. Furthermore (as 
is easily checked given that we have a maximum weight representation), the 
repeated action of gravity generators move the density distribution toward 
larger $N$; that is, gravity has a definite direction toward classical 
space-times. Combining this with the fact that classical space-times are the 
most
abundant ones ($D^{(N)}-D^{(N-1)}$ increases with $N$), we could explain why 
Universe geometry is almost classical. If this 
construction is considered as a model for gravity, it must be remembered that
one is not dealing 
with Einstein gravity, but rather with a higher-order correction to it
(probably more related to a Wess-Zumino-Witten-like gravity).
(Non-commutative) Einstein 
gravity should be studied in each of the hyperboloids that appear in the 
model,
by introducing a metric connection notion with a dynamical content. In two 
dimensions, classical Einstein gravity is trivial, and thus we have not 
concerned ourselves with it. However, in higher dimensions this problem 
should be faced. 
We stress that the model is not incompatible with, but rather
defines a framework to study, Einstein gravity.
 
As regards space-time operators, one must construct hermitian operators to
give an observable character to the position of a point. Thus,
\bea
x_{u}^{(N)}&=&x_{1}^{(N)}+x_{-1}^{(N)}\\
x_{v}^{(N)}&=&i(x_{1}^{(N)}-x_{-1}^{(N)}) . \nn
\eea
In this variables the $sl(2,{\mathbb R})$ Casimir constraint is given by:
\be
R^2 ={x_{u}^{N}}^2+{x_{v}^{N}}^2-{x_{0}^{N}}^2 . \label{ligar}
\ee
This expression does not distinguish between {\it de Sitter} and 
{\it anti-de Sitter} space-times, which in two dimensions are topologically 
identical. In fact,
the reconstruction of a geometry from a $C^*$-algebra does not provide a 
metric structure and thus such a distinction should not be expected at this 
level. There is freedom in choosing any of these by selecting an appropriate 
form of the $SL(2,{\mathbb R})$ Killing metric on the $(x_{u}^{N},x_{v}^{N},
x_{0}^{N})$
space, which induces the corresponding metric on the hyperboloid through 
(\ref{ligar}).


\begin{thebibliography}{99}

\bibitem{Madore} J. Madore, Class. Quant. Grav., {\bf 9} 69 (1992); 
                 J. Madore, Annals of Physics, {\bf 219} 187 (1992)


\bibitem{Polyakov} A.M.Polyakov, Mod. Phys. Lett. {\bf A11}, 893 (1987) 

\bibitem{Alekseev} A.Alekseev and S.Shatashvili, Nuc. Phys. {\bf B323}, 719
                   (1989)



\bibitem{Miguel}  V.Aldaya, J.Navarro-Salas and M.Navarro, Phys.Lett. 
                  {\bf B260}, 311 (1991)

\bibitem{Julio}  V.Aldaya, J.Guerrero and G.Marmo, in {\it Symmetries in Science X}, Ed. Bruno Gruber, Plenum Press (1998), pag. 1-36. 

\bibitem{Kac} V.G. Kac, Lecture Notes in Physics {\bf 94}, 441 (1979)

\bibitem{Feigin&Fuchs} B.L. Feigin and D.B. Fuchs, Funct. Anal. Appl. 
                        {\bf 16}, 114 (1982)

\bibitem{Thorn} C.B. Thorn, Nucl. Phys. {\bf B 248}, 551 (1984)

\bibitem{Senechal} P. Di Francesco, P. Mathieu, D. S\'en\'echal,
                {\it Conformal  field theory}, Springer-Verlag, 1997

\bibitem{Navarro-Salas} V.Aldaya, J.Navarro-Salas, Commun. Math. Phys. 
                        {\bf 126}, 575 (1990)\\
                        V.Aldaya, J.Navarro-Salas, Commun. Math. Phys.
                        {\bf 139}, 433 (1991) 

\bibitem{QG2} V. Aldaya, J.L. Jaramillo, Class. Quantum Grav. {\bf 17},
                1649 (2000).


\bibitem{Connes} Alain Connes, {\it Noncommutative geometry}, Academic Press,
                1994






\end{thebibliography}
\end{document}